\documentclass[twocolumn,showpacs,preprintnumbers,amsmath,amssymb]{revtex4}

\usepackage{graphicx}
\usepackage{dcolumn}
\usepackage{bm}

\newcommand{\invA}{\rm \AA^{-1}}

\begin{document}

\title{Anharmonic Phonons and Magnons in BiFeO$_3$}

\author{O.~Delaire$^1$}
\author{M.B. Stone$^1$}
\author{J. Ma$^1$ }
\author{A. Huq$^1$}
\author{D. Gout$^1$}
\author{C. Brown$^2$}
\author{K.F. Wang$^3$}
\author{Z.F. Ren$^3$}

\affiliation{ 1. Oak Ridge National Laboratory; Oak Ridge TN 37831  USA \\
2. NIST Center for Neutron Research; Gaithersburg MD 20899 USA\\
3. Department of Physics; Boston College; Boston MA 02467 USA \\ 
}

\date{\today}

\begin{abstract}
The phonon density of states (DOS) and magnetic excitation spectrum of polycrystalline BiFeO$_3$ were measured for temperatures $200 \leq T \leq 750\,$K , using inelastic neutron scattering (INS). Our results indicate that the magnetic spectrum of BiFeO$_3$ closely resembles that of similar Fe perovskites, such as LaFeO$_3$, despite the cycloid modulation in BiFeO$_3$. We do not find any evidence for a spin gap. A strong $T$-dependence of the phonon DOS  was found, with a marked broadening of the whole spectrum, providing evidence of strong anharmonicity. This anharmonicity is corroborated by  large-amplitude motions of Bi and O ions observed with neutron diffraction. A clear anomaly is seen in the $T$ dependence of Bi-dominated modes across the N\'{e}el transition. These results highlight the importance of spin-phonon coupling in this material.


\end{abstract}


\maketitle

\section{Introduction}

Multiferroic materials exhibiting a strong  magneto-electric coupling  are of great interest for potential applications in spintronic devices and actuator systems \cite{Wang-Liu-Ren, Catalan}. BiFeO$_3$ (BFO) is one of the few known systems exhibiting simultaneous magnetic and ferroelectric ordering at $T>300\,$K, and as such is a strong candidate for applications \cite{Wang-Liu-Ren, Catalan, Lebeugle-PRB2007}. BFO crystallizes in a rhombohedrally-distorted perovskite structure (space group $R3c$) \cite{Kubel, Palewicz-synchrotron, Palewicz-neutron, Sosnowska-review}. The Bi lone-pair is thought to be responsible for the off-centering of Bi atoms, which induces the ferroelectricity, with a high Curie temperature $T_{\rm C} \simeq 1100 \,$K \cite{Catalan}. The Fe ions, inside oxygen octahedra, carry large magnetic moments $\simeq 4 \mu_{\rm B}$ \cite{Catalan}, and order antiferromagnetically (AF) below the N\'{e}el temperature, $T_{\rm N} \simeq 640 \,$K, with some canting of the spins, and a long-period  cycloid modulation \cite{Catalan, Sosnowska, Ramazanoglu}. Unravelling the behavior of phonons and magnons, and their interactions, is crucial to understanding and controlling multiferroic properties \cite{Wang-Liu-Ren}. Phonons couple to the ferroelectric order, and magnons to the magnetic order, and it is expected that phonons and magnons strongly interact in a system exhibiting simultaneous ferroelectric and antiferromagnetic order, such as BFO \cite{Wang-Liu-Ren}. This interaction also gives rise to mixing of the excitations, resulting in electromagnons, for example \cite{Wang-Liu-Ren, Catalan}.  To our knowledge, there are currently no reported experimental data of the full magnon and phonon spectra in BFO, however. Multiple Raman measurements have been performed, but these only probe modes at small wavevectors $q \rightarrow 0$ \cite{Haumont-Raman, Cazayous-Raman, Rovillain-Raman, Singh2008-Raman, Singh2011-Raman, Shimizu-Raman, Hlinka-Raman, Porporati-Raman, Fukumura-Raman-lowT, Fukumura-Raman-highT, Palai-Raman}. Here, we report the first INS measurement of the phonon DOS in polycrystalline BiFeO$_3$,  as a function of $T$, as well as more detailed measurements of the magnon spectrum than previously reported \cite{Loewenhaupt}.  From these data, we  extract the exchange coupling constant of BFO, and we identify a strong anharmonicity of the phonons, providing evidence for strong spin-phonon coupling.

\section{Neutron Diffraction}

A stoichiometric mixture of Bi$_2$O$_3$ (99.99\%, Aldrich) and Fe$_2$O$_3$ (99.99\%, Aldrich) was ball-milled for 10 hours \cite{NIST-waiver}. The resulting powder was hot-pressed at $900^{\circ}$C for $5\,$min in a half-inch diameter graphite die, with a $2\,$ton force applied, and a heating rate of $300^{\circ}$C$/$min. The pressed pellets were annealed in vacuum at $750\,$K for $24\,$h. The total sample weight was about $15\,$g. 

\begin{table*}
\caption{Results of Rietveld refinements of time-of-flight neutron diffraction data for BiFeO$_3$ ($R3c$).  \label{Table1}}
 \begin{ruledtabular}
 \begin{tabular}[c]{cccccccccccc}
  $T$ &  $\chi^2$ &   $x_{\rm Fe_{2}O_{3}}$ &    $a,b$ & $c$ & ${\rm rms}\,U_{{\rm Bi}\perp c}$ & ${\rm rms}\,U_{{\rm Bi} \parallel c}$ & ${\rm rms}\,U_{{\rm Fe}\perp c}$ & ${\rm rms}\,U_{{\rm Fe} \parallel c}$ & ${\rm rms}\,U_{{\rm O,long}}$ & ${\rm rms}\,U_{{\rm O, mid}}$ & ${\rm rms}\,U_{{\rm O, short}}$\\
   (K)  & {}  & (\%)  & ($\AA$) & ($\AA$) & ($\AA$) & ($\AA$) & ($\AA$) & ($\AA$) & ($\AA$) & ($\AA$) & ($\AA$) \\
\hline
300 & 1.285 & 2.0 & 5.5752 & 13.8654 & 0.0977 & 0.0819 & 0.0694 & 0.0722 & 0.1038 & 0.0899 & 0.0698 \\
473 & 2.306 & 2.0 & 5.5862 & 13.9013 & 0.1291 & 0.1035 & 0.0914 & 0.0919 & 0.1303 & 0.1148 & 0.0956 \\ 
573 & 2.151 & 1.9 & 5.5930 & 13.9222 & 0.1455 & 0.1163 & 0.1008 & 0.1014 & 0.1458 & 0.1261 & 0.1058 \\
623 & 2.099 & 2.0 & 5.5966 & 13.9332 & 0.1527 & 0.1189 & 0.1046 & 0.1088 & 0.1531 & 0.1327 & 0.1094 \\
723 & 1.714 & 1.7 & 5.6033 & 13.9513 & 0.1645 & 0.1301 & 0.1153 & 0.1170 & 0.1658 & 0.1437 & 0.1189 \\
773 & 1.749 & 1.7 & 5.6065 & 13.9594 & 0.1718 & 0.1353 & 0.1190 & 0.1225 & 0.1704 & 0.1493 & 0.1264
\end{tabular}
\end{ruledtabular}
\end{table*}

\begin{figure}
\begin{center}
\includegraphics[width = 8.5 cm]{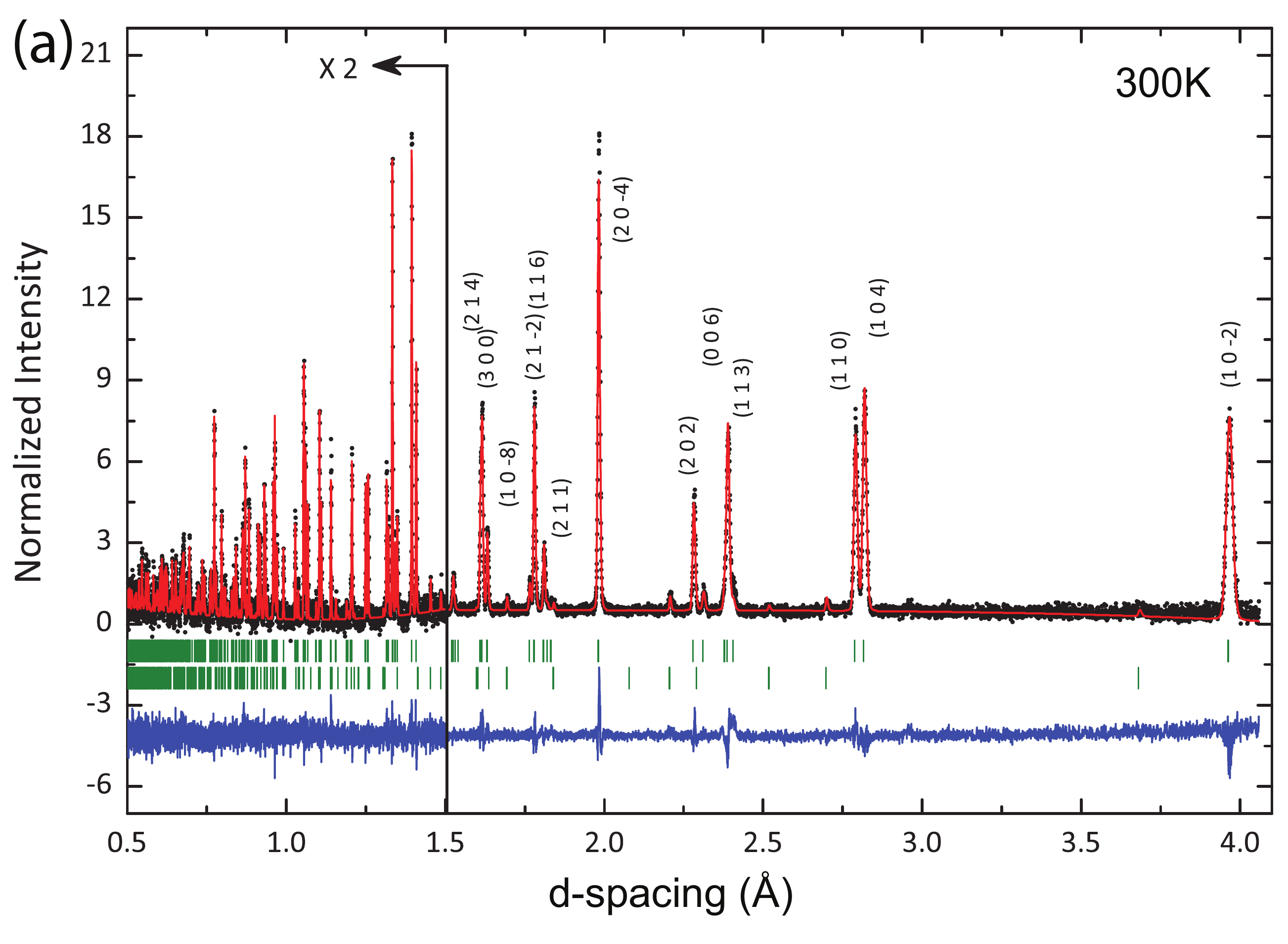}
\includegraphics[width = 8.5 cm]{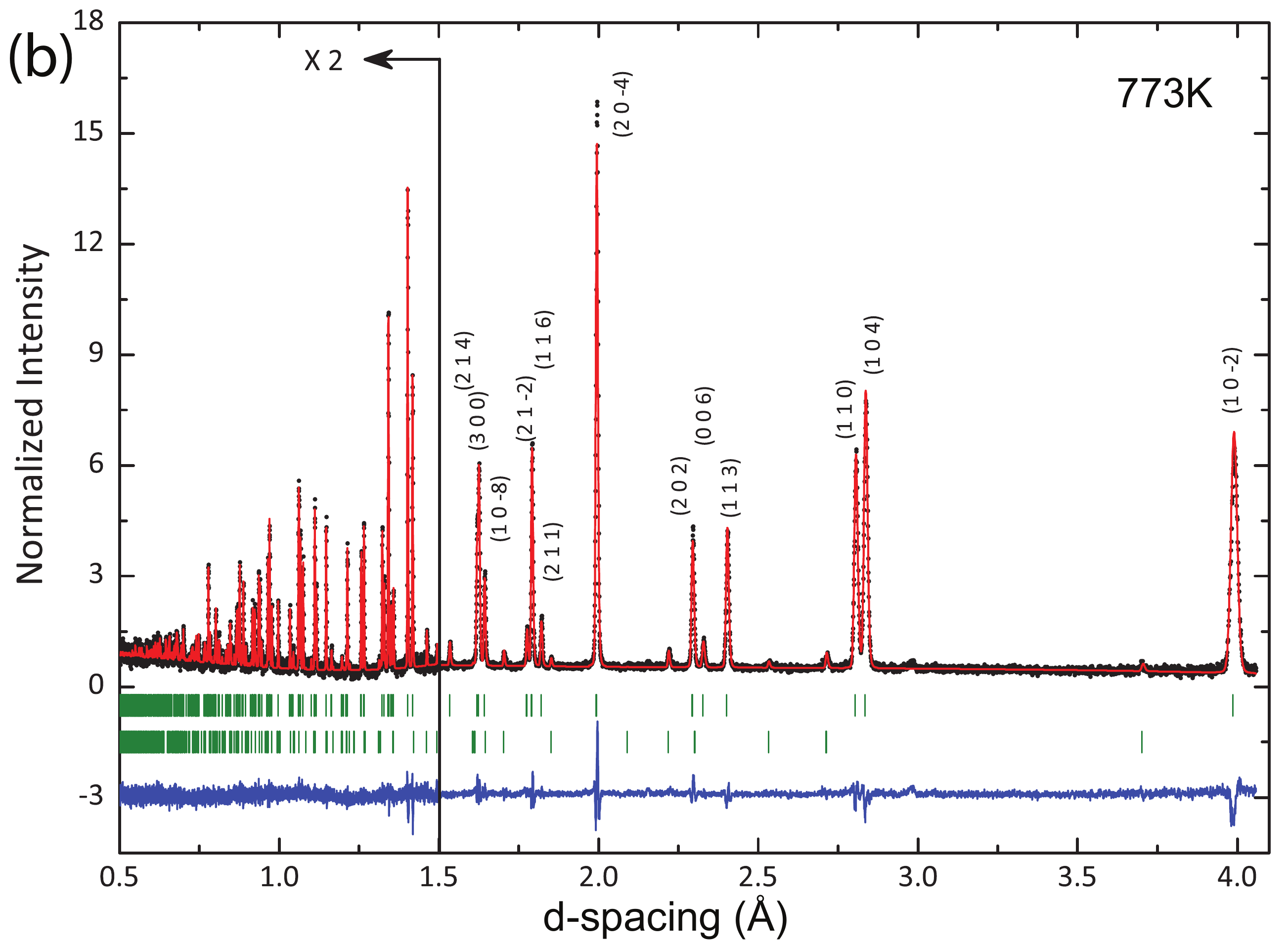}
\caption{Neutron dffraction patterns (markers) and Rietveld refinements (red line) for BiFeO$_3$ at $300\,$K (a) and $773\,$K (b).  Blues curves are the bottom of each panel are difference curves. Upper and lower tick marks are the reflection positions for the BiFeO$_3$ and Fe$_2$O$_3$ phases, respectively. Peak labels are for the BiFeO$_3$ phase. The refinements were done with the $R3c$ space-group. The data for $T<640\,$K were refined with a G-type antiferromagnetic structure (without cycloid modulation). } 
\label{powgen_fits}
\end{center}
\end{figure}

Neutron diffraction measurements were performed using the POWGEN time-of-flight diffractometer at the Spallation Neutron Source (SNS), at Oak Ridge National Laboratory. The sample was placed inside a thin-wall vanadium container, and loaded in a radiative vacuum furnace. Data were collected at $\rm T=300, 473, 573, 623, 723, 773\,$K. The time-of-flight diffractometer uses a broad incident spectrum of neutrons, and was configured with a center-wavelength $\lambda=1.066\,\AA$. The diffraction data were refined with GSAS \cite{Larson2004}, using the $R3c$ space-group. The fits indicated good crystallinity and good sample purity with $\sim$98\% BiFeO$_3$ and $\sim$2\% of a secondary phase, indexed as Fe$_2$O$_3$, at all temperatures.  For $T<640\,$K, the data were refined with a G-type antiferromagnetic order, without cycloid modulation. Since the inclusion of the G-type AF order did not change the refinements significantly, it is likely that the further incorporation of the cycloid structure would only have a minimal effect on our results. Results are summarized in Table~\ref{Table1}. The diffraction data and Rietveld fits for $T=300\,$K and $773\,$K are shown in Fig.~\ref{powgen_fits} (for one of three detector banks). We observe a constant intensity ratio for (104) and (110) reflections at all temperatures measured, and thus our data do not support the $R3c$--$R3m$ transition reported in Ref.~\cite{Jeong}.

\begin{figure}
\begin{center}
\includegraphics[width = 8.0 cm]{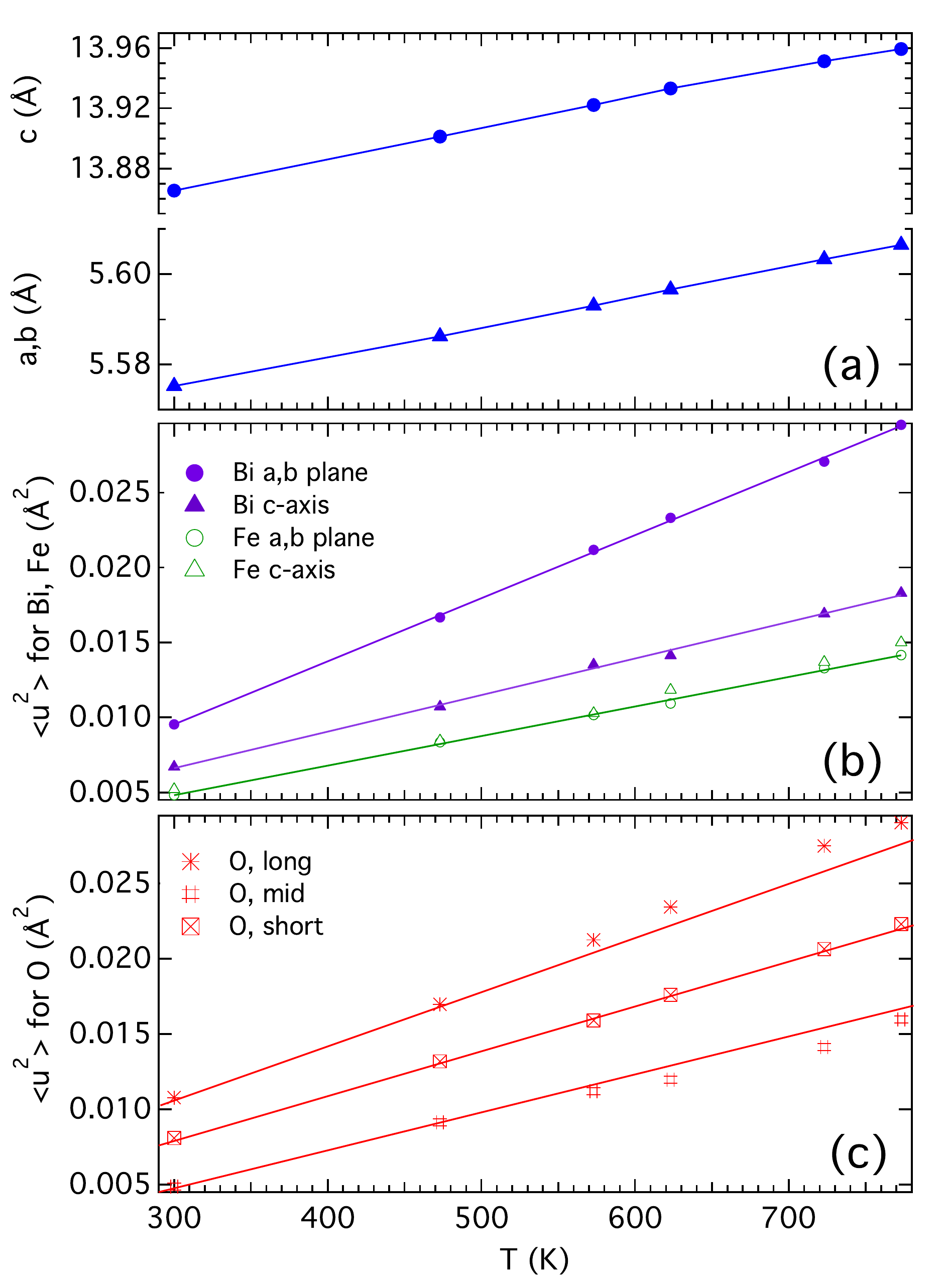}
\caption{Results of Rietveld refinements (space-group $R3c$) for lattice parameters and anisotropic atomic mean-square displacements, from POWGEN neutron diffraction data. Straight lines in (b) are fits to the data, while straight lines in (c) are guides to the eye.} 
\label{powgen_ADPs}
\end{center}
\end{figure}

The refined lattice parameters and atomic positions are in good agreement with prior reports \cite{Palewicz-synchrotron, Palewicz-neutron}. The temperature dependences of the lattice constants and  mean square thermal displacements (squares of quantities in Table~\ref{Table1}) are shown in Fig.~\ref{powgen_ADPs}. The mean-square displacements were refined with an anisotropic harmonic model for all atoms, which assumes Gaussian probability distributions for atom positions. The atomic mean-square displacements are largest for Bi, followed by O.  Our results for thermal displacements are also in good agreement with Ref.~\cite{Palewicz-neutron}, although we find displacements that are larger for Bi than for Fe, both along the (trigonal) $c$-axis and in the (basal) $a,b$ plane. For Bi and Fe vibration modes, the behavior of $\langle u^2 \rangle$ is linear in $T$, as expected in the high-$T$ regime of an harmonic oscillator (from the theoretical partial phonon DOS reported in \cite{Wang-DFT}, the average phonon energies for Bi and Fe vibrations are $12$ and $28\,$meV, respectively equivalent to $140\,$K and $320\,$K). In this regime, the amplitude of vibrations does not depend on the mass, but only on an effective force-constant, $K$, $\langle u^2 \rangle \propto T/K$ \cite{Willis-Pryor}. Linear fits to the Bi and Fe data indicate that the effective $K$ for Bi motions in the ($a,b$) plane is about half of the force-constant for Bi motions along $c$, which itself is comparable to those for either types of Fe motions. We note that our fits did not use anharmonic displacement models, which could slightly affect the results. The results for $\langle u^2 \rangle$ of oxygen atoms in Fig.~\ref{powgen_ADPs}-c shows a departure from linearity around $600\,$K for the short and long axes of the thermal ellipsoids. This effect may be related to the magnetic transition at $T_{\rm N} = 640 \,$K, although there could also be effects of phonon thermal occupations, since the average energy of O vibrations is $45\,$meV, corresponding to $520\,$K \cite{Wang-DFT}.

\section{Inelastic Neutron Scattering}

INS spectra were measured using the ARCS direct-geometry time-of-flight chopper spectrometer at the SNS \cite{arcspaper}. 
In the ARCS measurements, the sample was encased in a $12\,$mm diameter, thin-walled Al can, and mounted inside a low-background  furnace for measurements at $T=300, 470, 570, 690, 750\,$K. All measurements were performed under high vacuum. An incident neutron energy $E_{i}=110\,$meV was used and the energy resolution (full width at half max.) was $\sim$3$\,$meV at $80\,$meV neutron energy loss, increasing to $\sim$7$\,$meV at the elastic line. 

Additional INS measurements were performed with the Disk Chopper Spectrometer (DCS) at the NIST Center for Neutron Research, with incident energy $E_i = 3.55\,$meV, in up-scattering mode (excitation annihilation) \cite{dcspaper}. In these conditions, the energy resolution was $\sim$0.12$\,$meV at the elastic line, increasing to $\sim$1.2$\,$meV at 25$\,$meV neutron energy gain. For DCS measurements, the sample was encased in the same Al can as in the ARCS measurements.  The empty Al sample container was measured in identical conditions to the sample at all temperatures, and subtracted from the data. DCS measurements at $T=200, 300, 470, 570\,$K were performed in a high-temperature He refrigerator, and measurements at $T=570, 670, 720\,$K were performed in a radiative furnace. 

\subsection{Magnetic Spectrum}

\begin{figure}
\begin{center}
\includegraphics[width = 7.5 cm]{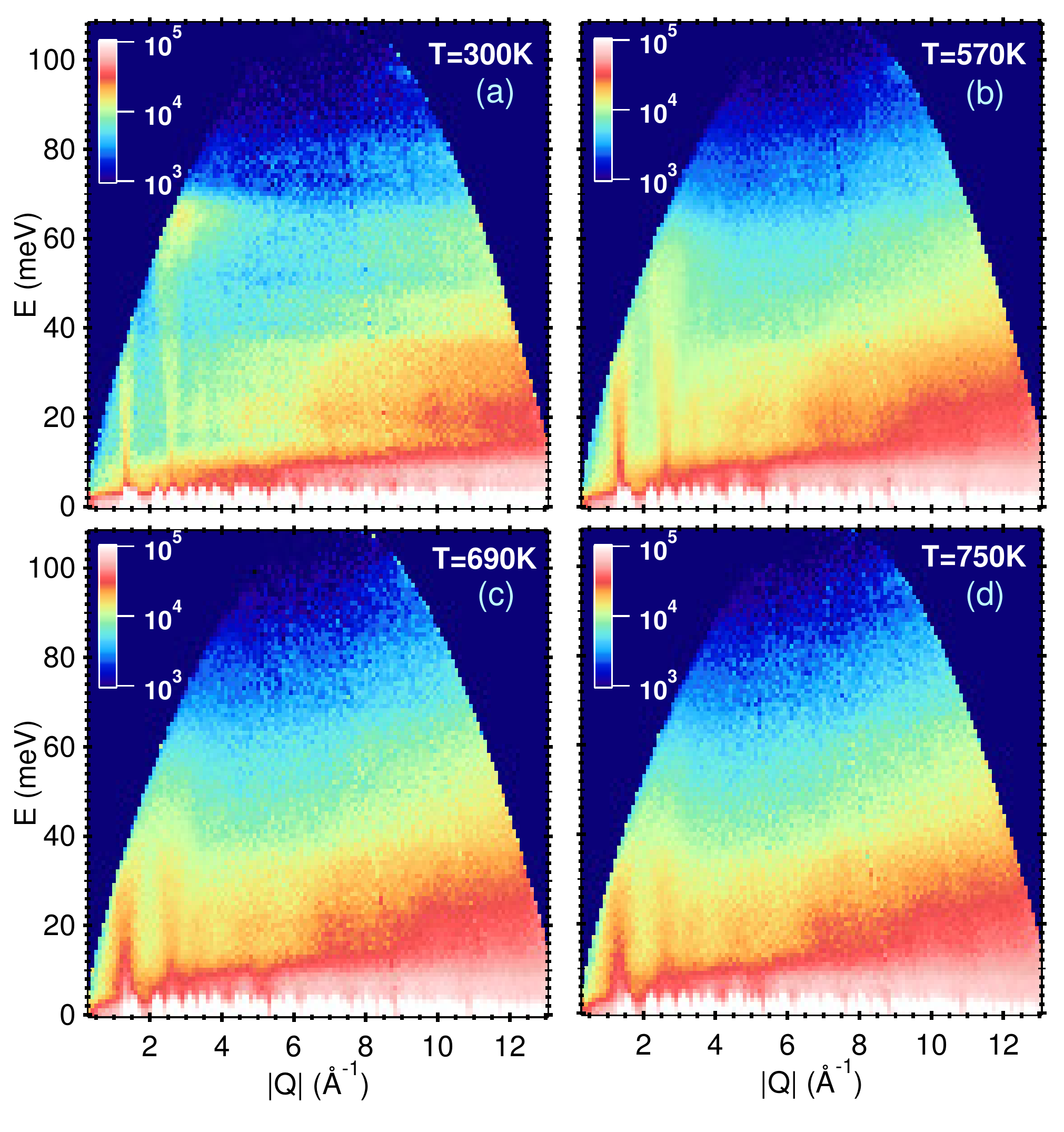}
\caption{$S(Q, E)$ for BiFeO$_3$ at different temperatures, measured using ARCS (logarithmic intensity).}
\label{ARCS_SQE_110meV}
\end{center}
\end{figure}

\begin{figure}
\begin{center}
\includegraphics[width = 7.5 cm]{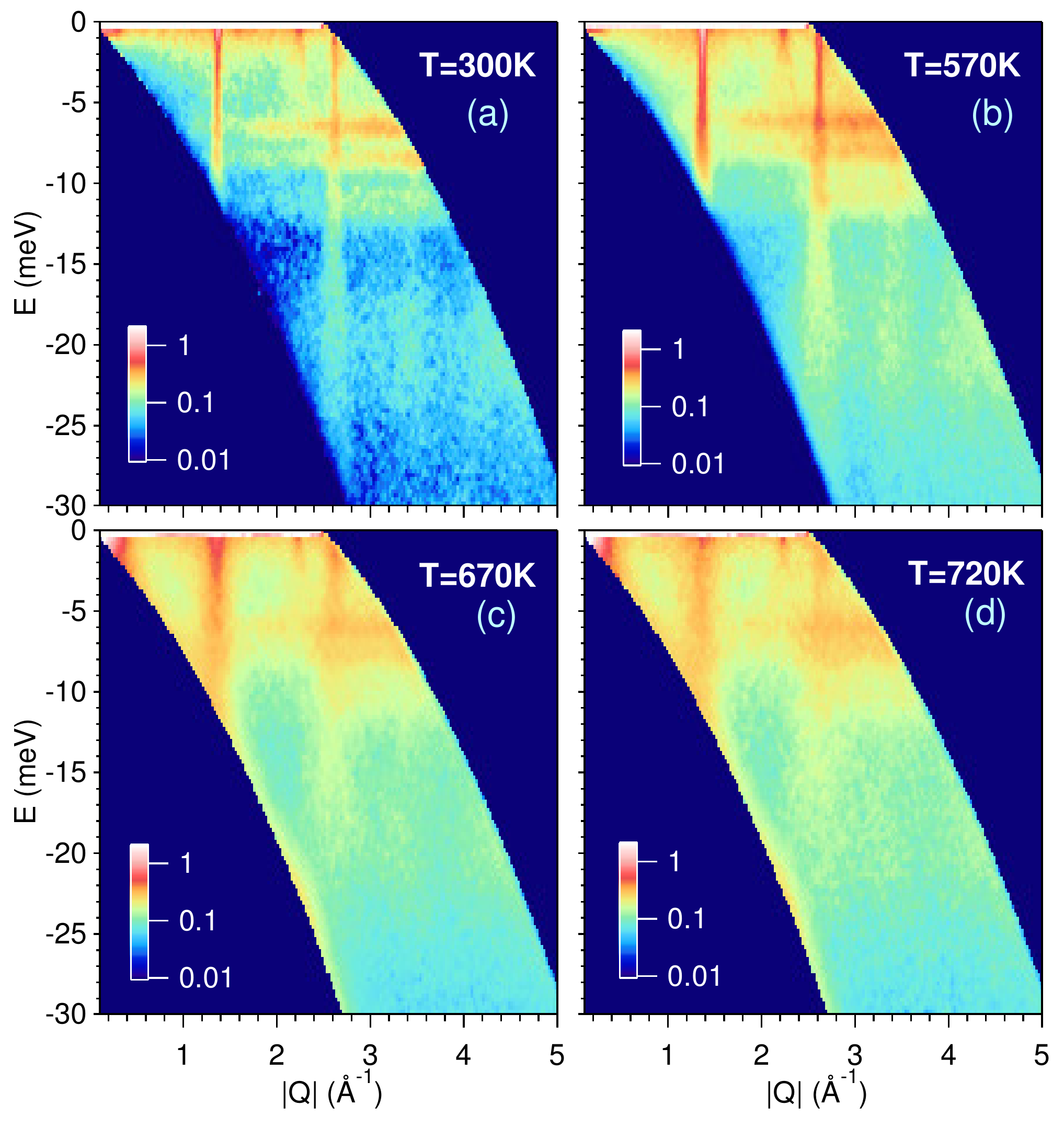}
\caption{ $S(Q,E)$ for BiFeO$_3$ at different temperatures, measured using DCS (logarithmic intensity).} 
\label{DCS_SQE}
\end{center}
\end{figure}

Figure~\ref{ARCS_SQE_110meV} shows the orientation-averaged scattering function, $S(Q,E)$, obtained with ARCS, as a function of temperature ($Q$ and $E$ are the momentum and energy transfer to the sample, respectively). At $T=300\,$K, the data for $Q<4 \invA$ clearly show steep spin-waves, emanating from strong magnetic Bragg peaks at $Q=1.37$ and $2.62\invA$, and extending to $E\sim70\,$meV. This range of energies overlaps with much of the phonon spectrum (see below). The intensity of the magnetic Bragg peaks decreases with increasing $T$, vanishing for $T\geq 670\,$K, in good agreement with the reported $T_{\rm N}=640\,$K. The magnetic scattering nearly vanishes for $Q>6\invA$, owing to the magnetic form factor of Fe$^{3+}$ ions. The high-$E$, optical part of the spin-waves is strongly damped for $T=470,570\,$K, well below $T_{\rm N}$. The low-$E$ part of the spin-waves is also clearly seen in the DCS data in Fig.~\ref{DCS_SQE}-a,b, corresponding to sharp vertical streaks emanating from the magnetic Bragg peaks. The acoustic part of the spin-waves shows some $Q$-broadening with increasing $T$ below $T_{\rm N}$. Magnetic correlations remain for $T>T_{\rm N}$, but are much broader (Figs.~\ref{ARCS_SQE_110meV}/\ref{DCS_SQE}-c,d). We note that the orientation-averaged spin-waves  show a strong similarity between  BiFeO$_3$ and other AFeO$_3$ perovskites, such as ErFeO$_3$, LaFeO$_3$ and YFeO$_3$ \cite{shapiro1974, McQueeney, JieMa_phd}.

\begin{figure}
\begin{center}
\includegraphics[width = 7.0 cm]{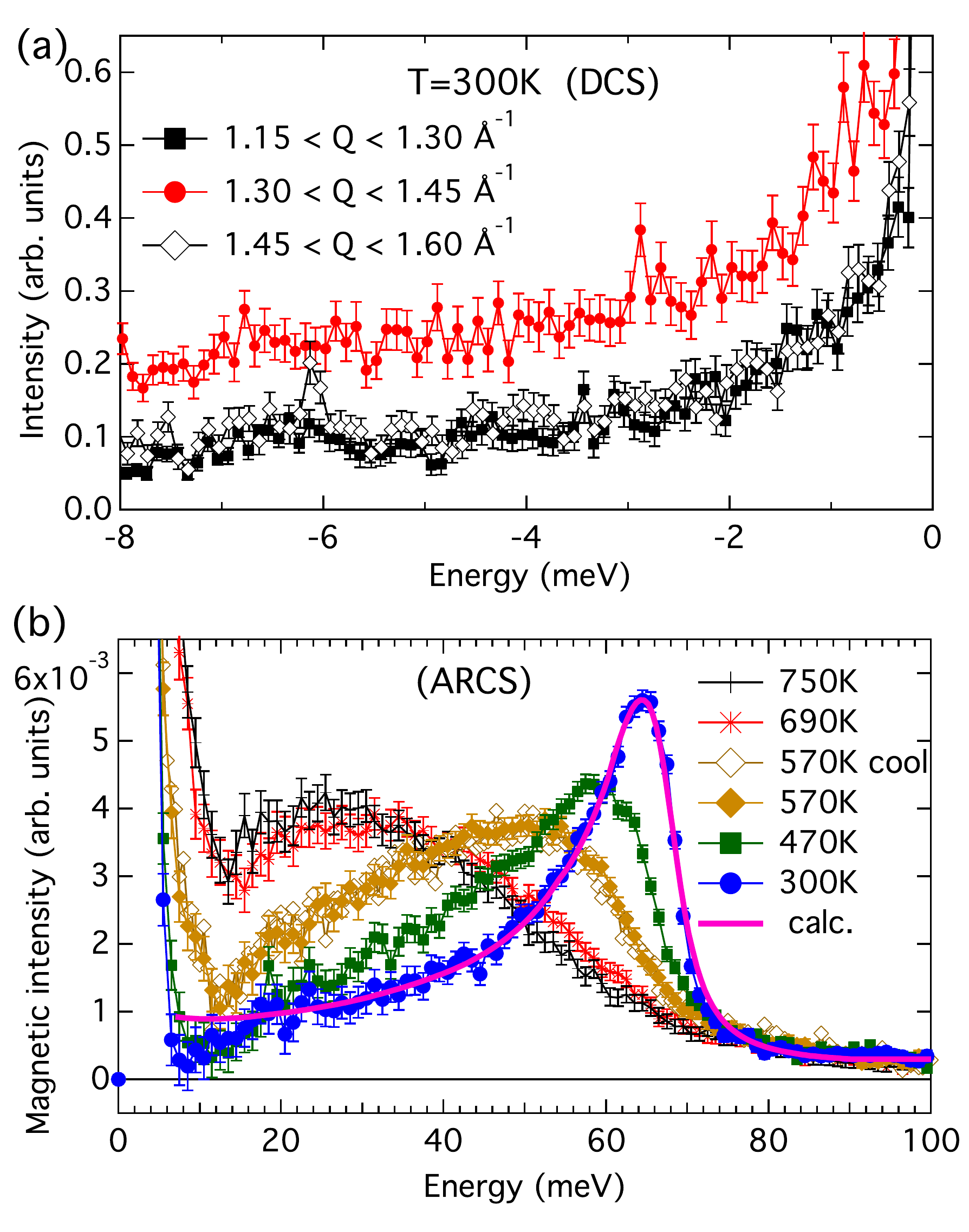}
\caption{(a) Integrated INS  intensity (DCS, $300\,$K) from spin-wave, $1.3\leqslant Q \leqslant1.45\,\invA$, compared with background phonon signal ($1.15\leqslant Q \leqslant1.3\,\invA$ and $1.45 \leqslant Q \leqslant 1.6\,\invA$). (b) Magnetic excitation spectra obtained from INS data (ARCS), integrated over $2 \leqslant Q \leqslant 6\, \invA$,  corrected for phonon scattering ($6 \leqslant Q \leqslant 10\, \invA$, scaled). The thick line is the magnon spectrum calculation (see text). Error bars indicate one standard deviation.} 
\label{magnetic_spectrum}
\end{center}
\end{figure}

An important question is whether low-energy spin-waves are present in BFO, and whether an energy gap exists, potentially associated with an anisotropy in the exchange coupling, or single-ion anisotropy. Figure~\ref{magnetic_spectrum}-b shows the magnetic scattering intensity, $S_{\rm mag}(E)$, measured with DCS, integrated over the spin-wave dispersing from $Q=1.37 \invA$, and compared to the phonon background on either side.  This figure clearly shows that the $S_{\rm mag}(E)$ intensity from the spin-wave persists down to $|E|\simeq 0.3\,$meV, where it merges with the elastic scattering signal. Thus, we can estimate an upper-bound for any magnetic anisotropy gap, $E_g < 0.3\,$meV, if any such gap exists. This is at odds with reports of a gap, $E_g=6\,$meV, derived from modeling the low-$T$ specific heat $C_P$ \cite{Lu-SpecificHeat}.

Although the powder average of excitations does not allow us to determine if multiple spin-wave branches exist within the magnon DOS, we can compare our results with recent Raman scattering studies that reported multiple electromagnon modes in BFO between 1.5 and 7.5 meV \cite{Rovillain-Raman, Singh2008-Raman}.  We do not see these excitations in the powder spectrum at either $200$ or $300\,$K (see Fig.~\ref{magnetic_spectrum}-a), although high-resolution INS measurements on single-crystals may be needed to observe such effects. Our results for the magnetic spectrum show a single maximum, around $65\,$meV at $300\,$K, see Fig.~\ref{magnetic_spectrum}-b. This is at odds with the magnetic spectrum reported in Ref.~\cite{Loewenhaupt}, which showed additional maxima around $30$ and $55\,$meV. However, the  extra peaks in \cite{Loewenhaupt} are likely due to peaks in the phonon DOS at these energies (see below).  We use the magnetic DOS to estimate the exchange coupling, with a simple spin-wave model for a collinear Heisenberg G-type antiferromagnet in an undistorted perovskite structure.  A similar model successfully described the spin waves in the orthoferrite compounds ErFeO$_3$, TmFeO$_3$ \cite{shapiro1974}, as well as LaFeO$_3$ and YFeO$_3$ \cite{McQueeney, JieMa_phd}, which have close magnetic structures. This does not capture possible effects from the spiral spin structure in BFO, but these are expected to be limited, owing to the long period of the modulation.  Within this model, two exchange constants, $J$ and $J^{'}$, describe a gapless spin-wave dispersion, with acoustic and optic modes that meet at the magnetic zone boundaries. This behavior can be seen in the large patch of scattering intensity in the ARCS $300\,$K data near 65 meV (Fig.~\ref{ARCS_SQE_110meV}-a).  The $J$[$J^{'}$] exchange constant corresponds to coupled moments with spin-spin distances of 3.968[5.613$\pm0.025$] {\AA}. Assuming $S=5/2$ Fe$^{3+}$ moments, and comparing the maximum energy of the measured  and model spin-wave spectra, we are able to place limits on the values of $J$ and $J^{\prime}$.  We find that $J = 1.6^{+0.4}_{-0.2}$ and $J^{\prime} = -0.25^{+0.07}_{-0.17}$ where there is a linear dependence on these parameters within these bounds $J^{\prime} = 0.90(2) - J0.40(1)$.  The calculated magnetic DOS for $J=1.6$ and $J^{\prime}=-0.253$~meV agrees well with the phonon subtracted $T=300$~K data shown in Fig.~\ref{magnetic_spectrum}-b.

There is a clear softening and broadening of the spin-wave spectrum with increasing temperature from $300\,$K to $570\,$K.  Magnon-magnon and magnon-phonon interactions are likely both responsible for this softening \cite{Singh2008-Raman, lovesey_vol2}. It is possible that the strong softening of $S_{\rm mag}(E)$ in this range is related to the an anomalous magnetization below $T_{\rm N}$ \cite{Lu-SpecificHeat}. Above $T_{\rm N}$, we observe Lorentzian scattering intensity centered at $E=0\,$meV, typical of paramagnetic behavior (the dip below $20\,$meV is a result of imperfect phonon subtraction). 

\subsection{Phonons}

In both Figs.~\ref{ARCS_SQE_110meV} and \ref{DCS_SQE}, the horizontal bands of intensity increasing as $Q^2$ are orientation-averaged phonon dispersions. In the DCS data, three main horizontal bands are clearly observed at $|E|\simeq 6.5, 8, 11\,$meV,  corresponding to the top of acoustic phonon branches and low-$E$ optical branches, which mainly involve Bi vibrations (see below).  The acoustic branches are seen dispersing out of a nuclear Bragg peak at $Q\simeq2.25 \invA$. The phonon cutoff corresponding to the top of oxygen-dominated optic branches is seen at $E\simeq85\,$meV in Fig.~\ref{ARCS_SQE_110meV}-a. A strong broadening of the phonon modes with increasing temperature can be seen in both Figs.~\ref{ARCS_SQE_110meV} and \ref{DCS_SQE}. This broadening indicates a strong damping of phonons, over the full spectrum. The $T$ range over which this occurs is in agreement with prior Raman measurements \cite{Singh2011-Raman}, and points to a spin-phonon coupling effect.

\begin{figure}
\begin{center}
\includegraphics[width = 7.0 cm]{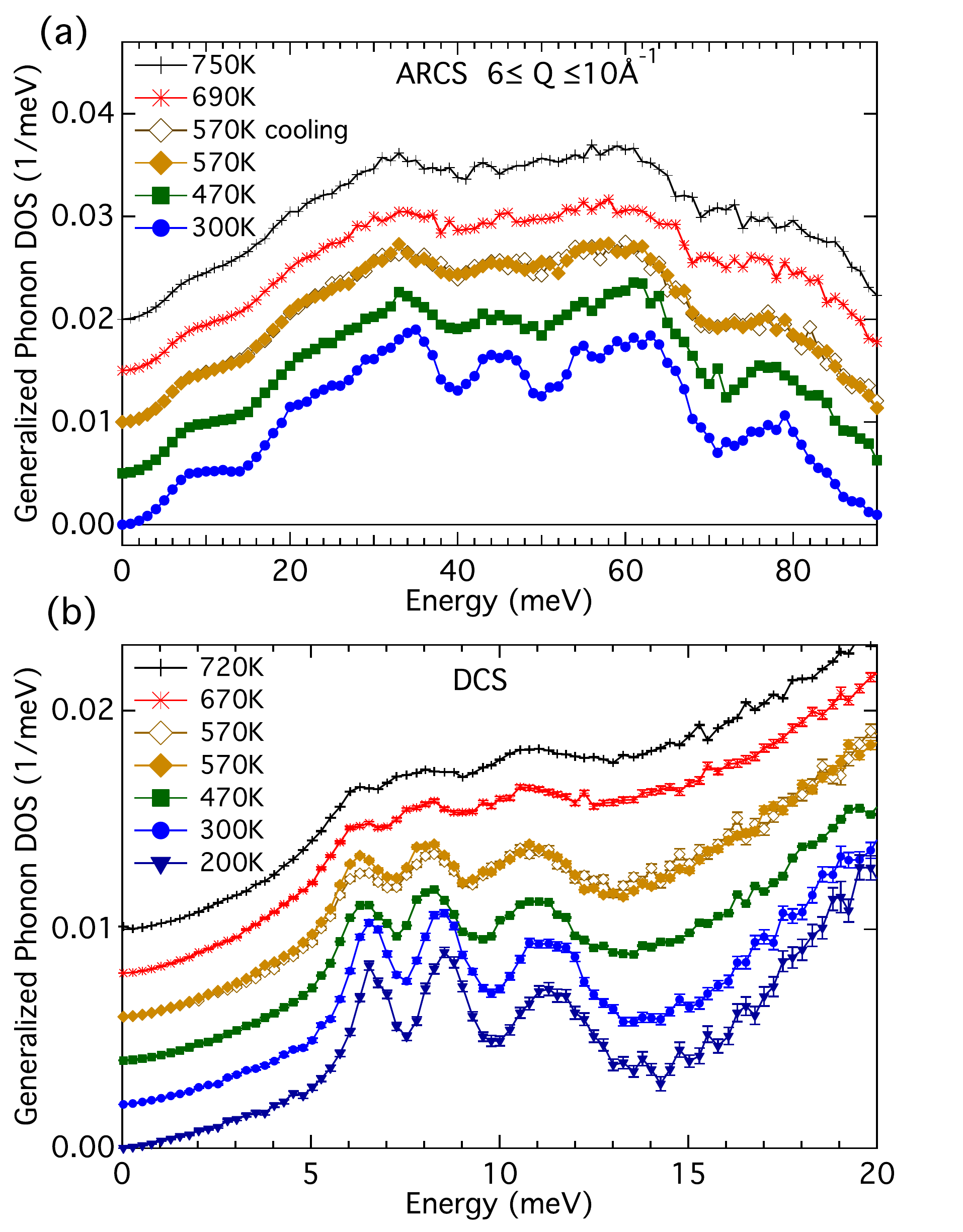}
\caption{(a) Generalized phonon DOS obtained from ARCS data. (b) Low-$E$ part of generalized DOS from DCS data, showing strong broadening of Bi-dominated modes. Curves for different temperatures are vertically offset for clarity.} 
\label{PhononDOS}
\end{center}
\end{figure}

\begin{figure}
\begin{center}
\includegraphics[width = 8.0 cm]{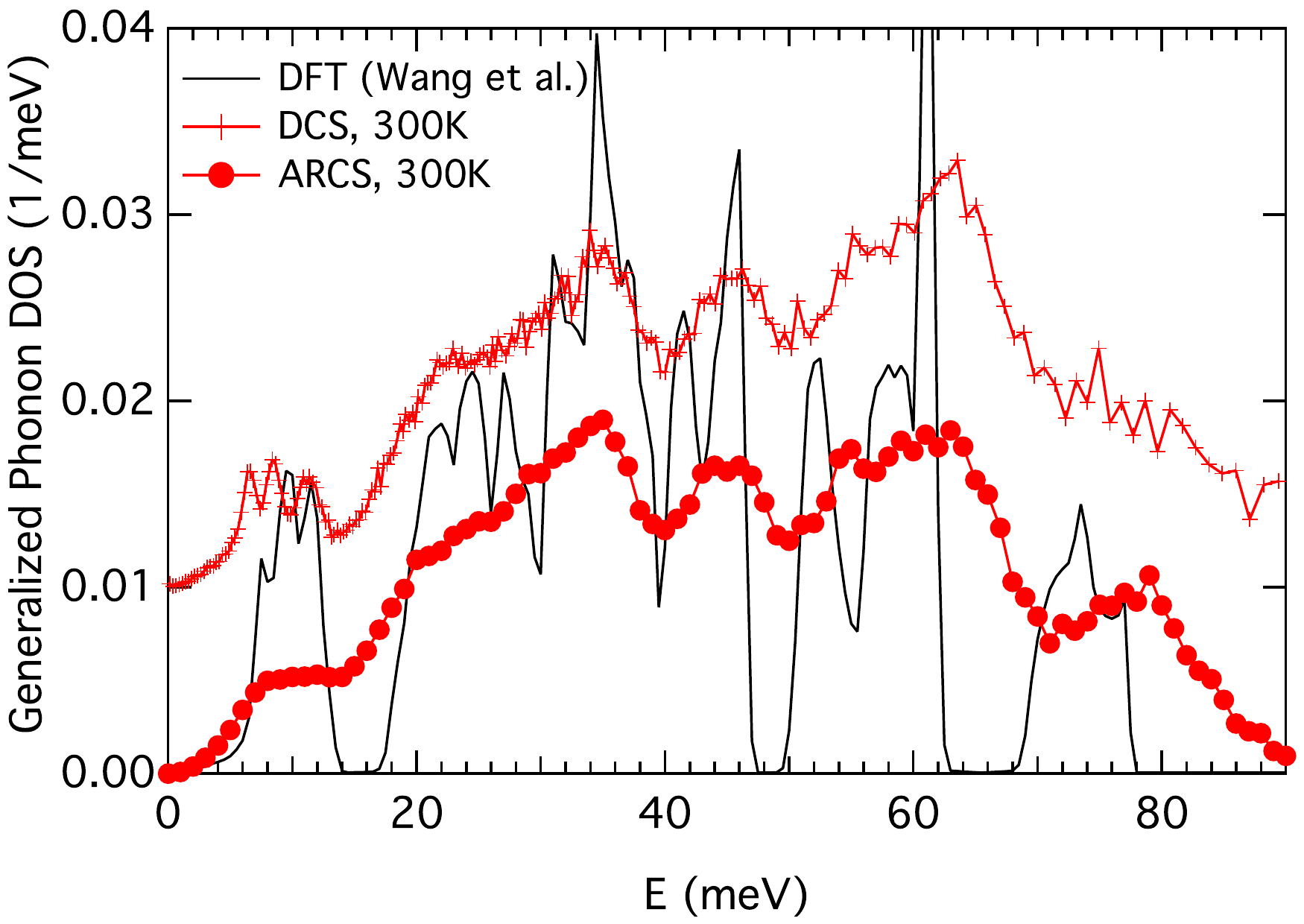}
\caption{Generalized phonon DOS of BiFeO$_3$ measured with INS (ARCS and DCS) at 300$\,$K, compared with first-principles calculation of Wang {\textit et al.}~\cite{Wang-DFT}. } 
\label{phononDOS_DFT_comparison}
\end{center}
\end{figure}

The $S(Q,E)$ data were analyzed to extract the generalized phonon DOS, $g(E)$, in the incoherent scattering approximation. 
The data from ARCS were integrated over $6\leqslant Q \leqslant 10 \invA$, which minimizes any contribution from magnetic scattering, and the elastic peak was subtracted, and extrapolated with a quadratic $E$ dependence for $E < 5\,$meV. A correction for multiphonon scattering was performed at all $T$ \cite{DANSE-ref, Kresch-Ni}. The gDOS from the DCS data was obtained by integrating over the full range of $Q$ (no multiphonon correction) \cite{dave}. The measured signal from the empty container was much weaker than from the sample, and was easily subtracted. The results are shown in Fig.~\ref{PhononDOS}(a) for the full gDOS (ARCS) and panel (b) for $E<20\,$meV (DCS). Although the DCS data include a magnetic contribution, this effect is limited, and the DOSs form ARCS and DCS are in excellent agreement, considering the difference in instrument resolution (see Fig.~\ref{phononDOS_DFT_comparison}). The coarser energy resolution in ARCS data washes out the three Bi-dominated peaks at $E \leqslant 15\,$meV, but the two DOS curves are otherwise very similar. In BFO, the different elements have different ratios of cross-section over mass, $\sigma / M$, resulting in a weighted phonon DOS (gDOS). The values of $\sigma / M$ for (Bi, Fe, O) are (0.044, 0.208, 0.265), in units of barns/amu, respectively. Thus, the modes involving primarily Bi motions are under-emphasized in $S(Q,E)$ and $g(E)$ (but there is relatively little weighting of Fe modes compared to O modes). 

While the energy-range of the spectrum is comparable with other Fe perovskites \cite{JieMa_phd}, what is striking is the severe broadening of the spectrum with increasing $T$. The  gDOS measured at $300\,$K is in good agreement overall with the first-principles calculation of Wang \textit{et al.} within spin-polarized DFT+U \cite{Wang-DFT}, as can be seen in Fig.~\ref{phononDOS_DFT_comparison}. This agreement allows for a clear identification of the main features in the DOS. The  three  peaks at $E\simeq6.5, 8, 11\,$meV (Fig.~\ref{PhononDOS}(b)) arise from the top of acoustic branches and the lowest-$E$ optic modes, and they are dominated by Bi motions. The lower two Bi peaks ($6.5\,$meV and $8\,$meV) are softer than predicted by DFT by about 10\% \cite{Wang-DFT}. These Bi modes are responsible for the peak in $C_P/T^3$ around $25\,$K reported in Ref. \cite{Lu-SpecificHeat}. The phonon spectrum for $E>40\,$meV is mainly comprised of oxygen vibrations, and is stiffer in the measured DOS than in the DFT calculation \cite{Wang-DFT}.  While the agreement is generally good between the DFT calculation of Wang \textit{et al.} and the phonon DOS measured at low temperature, the $T$ dependence of the DOS is strongly affected by anharmonicity and spin-phonon coupling, as we discuss in the next section.

\section{Discussion}

\begin{figure}
\begin{center}
\includegraphics[width = 6.5 cm]{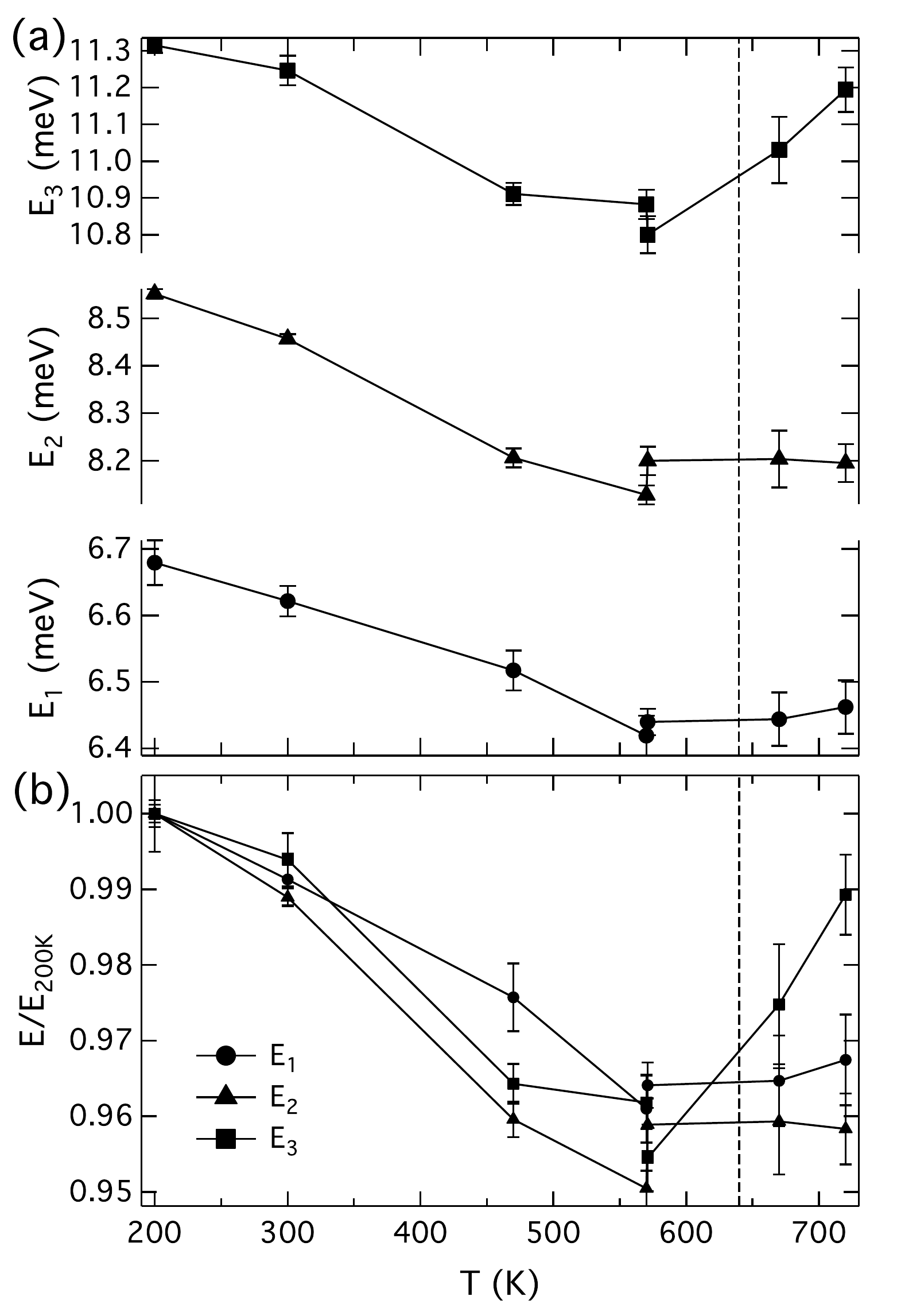}
\caption{(a) Centroids of the three low-energy peaks in DOS around $6.5$, $8.5$, $11\,$meV (resp. $E_1$, $E_2$, $E_3$), as a function of temperature. (b) Relative change in these energies with respect to their value at $200\,$K. The vertical dashed line at $T=640\,$K denotes the N\'{e}el temperature.} 
\label{softening}
\end{center}
\end{figure}

As already pointed out above, the phonon DOS exhibits a severe broadening with increasing $T$, which affects the whole spectrum. In particular, the broadening of Bi modes at low-$E$ is obvious in the DCS data, shown in Fig.~\ref{PhononDOS}-b. It is particularly strong between $470\,$K and $570\,$K, with little additional broadening observed further above $T_{\rm N}$. This  indicates a coupling between anharmonicity and the loss of the AF order. These results are consistent with the reported observations of strong broadening of Raman phonon modes \cite{Singh2011-Raman}. The oxygen modes, dominating the DOS between $40\,$meV and $85\,$meV, are also strongly broadened in this $T$-range. 

We now analyze in more detail the temperature dependence of the three peaks at $6.5$, $8.5$, $11\,$meV (resp. $E_1$, $E_2$, $E_3$). The peaks were fitted with Gaussians, and the resulting peak centers are plotted as a function of temperature in Fig.~\ref{softening}. As may be seen on this figure, all three peaks undergo a pronounced but gradual softening between $200\,$K and $570\,$K, besides the strong broadening. However, this softening stops around $T_{\rm N}$. This is compatible with the recently reported behavior of $A_1$ Raman modes \cite{Singh2011-Raman}. $E_3$ actually appears to stiffen above $T_{\rm N}$, but this may partly be due to the influence of  broadening and softening of modes at $E \geqslant 13\,$meV, causing a skewed contribution to the $E_3$ peak. 

The change in phonon softening behavior around $T_{\rm N}$ is further evidence of the influence of spin-phonon coupling. The softening of phonon modes with increasing $T$, as observed here for $T \leqslant 570\,$K, can generally be related to the thermal expansion of the system through the quasiharmonic approximation $d \ln E = - \gamma d \ln V$, with $\gamma$ the Gr\"{u}neisen parameter \cite{Grimvall-TPM99}. We determined the relative change in volume to be about 1.4\% between $200\,$K and $570\,$K from our diffraction measurements (this was linearly extrapolated over the range $200 - 570\,$K, since our diffraction data were limited to $T \geqslant 300\,$K). The average relative decrease of $E_1$, $E_2$, and $E_3$ is $-4.2\pm0.7$\% over the same $T$ range, yielding a Gr\"{u}neisen parameter $\gamma=3.0\pm0.5$. Such a large value of $\gamma$ is a further corroboration of the anharmonicity of these Bi-dominated modes. The lack of softening above $570\,$K indicates that an increase in interatomic force-constants associated with the loss of magnetic order compensates for the effect of thermal expansion.

We note that Bi and O atoms undergo large amplitude vibrations, according to both our measurements and reports of others \cite{Palewicz-neutron, Palewicz-synchrotron}. Since the Bi and O modes are sharp at $T\leqslant300\,$K, the large displacements observed in diffraction data are dynamic in nature, rather than associated with static disorder. These large amplitudes of vibration for Bi and O are related to the anharmonic scattering of phonons, which leads to the broadening and softening of features in the DOS. We have also performed measurements of the Fe-partial phonon DOS with nuclear-resonant inelastic x-ray scattering (NRIXS) on ${}^{57}$Fe-enriched samples, and observed a more limited broadening of Fe modes, in agreement with the smaller thermal displacements of these atoms \cite{Delaire-BFO-NRIXS}. 

We suggest that the large thermal displacements and anharmonicity of Bi and O modes lead to structural fluctuations, such as variations in Fe-O-Fe bond lengths and bond angles (tuning the superexchange interaction) through tilts and rotations of FeO$_6$ octahedra, that could lead to fluctuations in magnetic coupling. The magnitude of O thermal motions perpendicular to Fe-O bonds actually leads to fluctuations in the Fe-O-Fe bond angle that are larger ($\simeq 6-10^{\circ}$) than the variation of average angle with $T$. Reciprocally, the loss of magnetic order induces a change in interatomic force-constants, stiffening the Bi vibration modes at low $E$.  The motion of Bi atoms is also directly related to the ferroelectricity. The large-amplitude structural fluctuations could thus lead to steric effects between Bi motions and the rotations of oxygen octahedra, yielding a complex coupling between ferroelectric and AF magnetic orders.

\section{Summary}

We have systematically investigated the temperature dependence of the magnetic excitation spectrum and phonon density of states of BiFeO$_3$ over the range $200 \leqslant T \leqslant 750K$, using inelastic neutron scattering. In addition, we performed neutron diffraction measurements and refined the lattice parameters and thermal displacement parameters over $300 \leqslant T \leqslant 770K$. We separated the magnon and phonon contributions in the $S(Q,E)$ data, and observed a strong resemblance of the magnon spectrum with that of related compounds LaFeO$_3$ and YFeO$_3$. The magnon spectrum was fit satisfactorily with a simple collinear Heisenberg model for a G-type antiferromagnet,  indicating the limited role of the cycloid on the spin dynamics, as expected from the long period of the cycloid modulation. The phonon DOS obtained from the high-$Q$ part of the $S(Q,E)$ data is in good agreement at low temperatures with the first-principles calculation of Wang \textit{et al.} \cite{Wang-DFT}. However, the phonon DOS shows a strong temperature dependence, with in particular a pronounced broadening. Also, both the softening and broadening of features in the DOS correlate with the loss of antiferromagnetic order around $T_{N}=640\,$K, indicating the presence of significant spin-phonon coupling, in agreement with recently reported Raman measurements \cite{Singh2011-Raman}. The potential influence of large atomic displacements on the modulation of the superexchange interaction, and the concomitant effect of the change in force-constants from the loss of magnetic order were pointed out.

\section{Acknowledgements}

The Research at Oak Ridge National Laboratory's Spallation Neutron Source was sponsored by the Scientific User Facilities Division, Office of Basic Energy Sciences, US DOE. This work utilized facilities supported in part by the National Science Foundation under Agreement No. DMR-0944772. The work performed at Boston College is funded by the US Department of Energy under contract number DOE DE-FG02-00ER45805 (ZFR).


\begin{references}

\bibitem{Wang-Liu-Ren} K.F. Wang, J.-M. Liu, and Z.F. Ren, Adv. Phys. {\bf 58}, 321 (2009).

\bibitem{Catalan} G. Catalan and J. F. Scott, Adv. Mater. {\bf 21}, 2463 (2009).

\bibitem{Lebeugle-PRB2007}D. Lebeugle, D. Colson, A. Forget, M. Viret, P. Bonville, J. F. Marucco, and S. Fusil, Phys. Rev. B {\bf 76}, 024116 (2007).

\bibitem{Kubel} F. Kubel and H. Schmid, Acta Cryst. {\bf B46}, 698 (1990).

\bibitem{Palewicz-synchrotron}A. Palewicz, T. Szumiata, R. Przenioslo, I. Sosnowska, I. Margiolaki, Solid State Commun. {\bf 140}, 359–363 (2006).

\bibitem{Palewicz-neutron}A. Palewicz, R. Przenioslo, I. Sosnowska, and A. W. Hewat,  Acta Cryst.  {\bf B63}, 537 (2007).

\bibitem{Sosnowska-review} I.M. Sosnowska, J. Microsc. {\bf 236}, 109 (2009).

\bibitem{Sosnowska}I Sosnowska, T P Neumaier and E Steichele, J. Phys. C: Solid State Phys. {\bf 15}, 4835 (1982).

\bibitem{Ramazanoglu} M. Ramazanoglu, W. Ratcliff II, Y. J. Choi, Seongsu Lee, S.-W. Cheong, and V. Kiryukhin, Phys. Rev. B {\bf 83}, 174434 (2011).

\bibitem{Haumont-Raman}R. Haumont, J. Kreisel, P. Bouvier, and F. Hippert, Phys. Rev. B {\bf 73}, 132101 (2006).

\bibitem{Cazayous-Raman} M. Cazayous, Y. Gallais,  A. Sacuto, R. de Sousa, D. Lebeugle and D. Colson, Phys. Rev. Lett. {\bf 101}, 037601 (2008).

\bibitem{Rovillain-Raman}P. Rovillain, M. Cazayous, Y. Gallais,  A. Sacuto, R. P. S. M. Lobo, D. Lebeugle and D. Colson, Phys. Rev. B {\bf 79}, 180411(R) (2009).

\bibitem{Singh2008-Raman} M. K. Singh, R. S. Katiyar, and J. F. Scott, J. Phys.:Condens. Matter {\bf 20}, 252203 (2008).

\bibitem{Singh2011-Raman} M. K. Singh and R. S. Katiyar, J. Appl. Phys. {\bf 109}, 07D916 (2011).

\bibitem{Shimizu-Raman}T. Shimizu, T. Era, H. Taniguchi, D. Fu, T. Taniyama, and M. Itoh, Ferroelectrics {\bf 403}, 187-190 (2010).

\bibitem{Hlinka-Raman}J. Hlinka, J. Pokorny, S. Karimi, and I. M. Reaney, Phys. Rev. B {\bf 83}, 020101(R) (2011).

\bibitem{Porporati-Raman}A.A. Porporati, K. Tsuji, M. Valant, A.-K. Axelssond and G. Pezzottia, J. Raman Spectrosc. {\bf 41}, 84–87 (2010).

\bibitem{Fukumura-Raman-lowT}H. Fukumura, S. Matsui, H. Harima, T. Takahashi, T. Itoh, K. Kisoda, M. Tamada, Y. Noguchi and M. Miyayama, J. Phys.: Condens. Matter {\bf 19}, 365224 (2007).

\bibitem{Fukumura-Raman-highT}H. Fukumura, H. Harima, K. Kisoda, M. Tamada, Y. Noguchi, M. Miyayama, J. Magn. Magn. Mater. {\bf 310}, e367 (2007).

\bibitem{Palai-Raman}R. Palai, J. F. Scott, and R. S. Katiyar, Phys. Rev. B {\bf 81}, 024115 (2010).

\bibitem{Loewenhaupt} M. Loewenhaupt, Physica B {\bf 163}, 479 (1990).


\bibitem{NIST-waiver} Identification of commercial equipment or products in the text is not intended to imply any recommendation or endorsement by the National Institute of Standards and Technology.

\bibitem{Larson2004}
A. C. Larson and R. B. VonDreele, Los Alamos National Laboratory Report LAUR 86-748 (2004).

\bibitem{Willis-Pryor} B.T.M. Willis and A.W. Pryor, ``Thermal Vibrations in Crystallography'', (Cambridge University Press, Cambridge, 1975).

\bibitem{Jeong}Y.K. Jeong, C.W. Bark, S. Ryu, J.-H. Lee and H. M. Jang, J.  Korean Phys. Soc. {\bf 58}, 817-820 (2011).

\bibitem{arcspaper}
D. L. Abernathy, M. B. Stone, M. J. Loguillo, M. S. Lucas, O. Delaire, X. Tang, J. Y. Y. Lin, and  B. Fultz, Rev. Sci. Inst. (in review) (2011).

\bibitem{dcspaper}
J.R.D. Copley and J.C. Cook, Chem. Phys. {\bf 292}, 477 (2003).

\bibitem{shapiro1974} S. M. Shapiro, J. D. Axe and J. P. Remeika, Phys. Rev. B {\bf 10}, 2014, (1974).

\bibitem{McQueeney} R.J. McQueeney, J.-Q. Yan, S. Chang, and J. Ma, Phys. Rev. B {\bf 78}, 184417 (2008).

\bibitem{JieMa_phd}J. Ma, Ph.D. thesis, Iowa State University (2010).

\bibitem{Lu-SpecificHeat}J. Lu, A. G\"unther, F. Schrettle, F. Mayr, S. Krohns, P. Lunkenheimer, A. Pimenov, V.D. Travkin, A.A. Mukhin, and A. Loidl, Eur. Phys. J. {\bf B 75}, 451–460 (2010).

\bibitem{lovesey_vol2} S. W. Lovesey, \textit{Theory of Neutron Scattering from Condensed Matter:  Volume 2} (Clarendon Press, Oxford, UK 1987).

\bibitem{DANSE-ref}DANSE software project. http://danse.us

\bibitem{Kresch-Ni}M.G. Kresch, O. Delaire, R. Stevens, J.Y.Y. Lin and B. Fultz,
Phys. Rev. B, {\bf 75}, 104301 (2007).

\bibitem{dave} R.T. Azuah, L.R. Kneller, Y. Qiu, P.L.W. Tregenna-Piggott, C.M. Brown, J.R.D. Copley, and R.M. Dimeo, J. Res. Natl. Inst. Stan. Technol. {\bf 114}, 341 (2009).



\bibitem{Wang-DFT}Y. Wang, J.E. Saal, P. Wu, J. Wang, S. Shang, Z.-K. Liu, L.-Q. Chen, Acta Materialia {\bf 59}, 4229–4234 (2011).

\bibitem{Grimvall-TPM99}G. Grimvall, \textit{Thermophysical Properties of Materials} (North Holland,  Amsterdam, 1999).

\bibitem{Delaire-BFO-NRIXS}O. Delaire {\textit et al., to be published.}

\end{references}
\end{document}